# Five Minutes of DDoS Brings down Tor: DDoS Attacks on the Tor Directory Protocol and Mitigations

Zhongtang Luo
Purdue University
West Lafayette, USA
luo401@purdue.edu

Jianting Zhang
Purdue University
West Lafayette, USA
zhan4674@purdue.edu

Akshat Neerati
Purdue University
West Lafayette, USA
aneerat@purdue.edu

Aniket Kate
Purdue University / Supra Research
West Lafayette, USA
aniket@purdue.edu

## Abstract

The Tor network offers network anonymity to its users by routing their traffic through a sequence of relays. A group of nine directory authorities maintains information about all available relay nodes using a distributed directory protocol. We observe that the current protocol makes a strong synchrony assumption, which makes it vulnerable to natural as well as adversarial non-synchronous communication scenarios over the Internet. In this paper, we show that it is possible to cause a failure in the Tor directory protocol by targeting a majority of the authorities for only five minutes using a well-executed distributed denial-of-service (DDoS) attack. We demonstrate this attack in a controlled environment and show that it is cost-effective for as little as $53.28 per month to disrupt the protocol and to effectively bring down the entire Tor network. To mitigate this problem, we consider the popular partial synchrony assumption that ensures protocol security even when the network delays are large and unknown initially. We design a new Tor directory protocol that leverages a standard partial-synchronous consensus protocol to solve this problem, while also proving its security. We have implemented a prototype in Rust, demonstrating comparable performance to the current protocol.

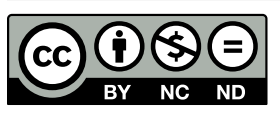

## 1 Introduction

The Tor network, a volunteer-driven anonymity service, caters to over 2 million users globally every day [37]. It allows users to route traffic through its relays, obscuring their IP and geographical metadata. This service is free, and the public can contribute by offering their servers as relays.

The Tor network allows the user to pick three of its relays and send its traffic through them in a chain. In this manner, the first relay cannot see the destination of the traffic, and the last relay cannot see the origin of the traffic, so no single relay can fully deanonymize the traffic. The efficacy and integrity of Tor hinge on clients possessing up-to-date relay information [43] from a large pool, as three relays controlled by the same party can easily track the user if it decides to route through them. Tor utilizes nine designated directory authorities to distribute relay information. To economize on bandwidth, rather than having each client request data from every authority individually, the authorities employ a distributed consensus protocol. Together, they sign on and distribute a unique version of the relay details called a *consensus document*. This ensures the data is both accurate and current, allowing any client to obtain and validate the information in a single step from any one of the authorities.

Although both clients and authorities benefit from this scheme of distributing relay information, the current directory protocol has been shown to be insecure to an attack. Recent research by Luo et al. [23] has illuminated vulnerabilities in the current directory protocol, revealing potential security breaches by malicious authorities. Furthermore, they proposed a secure protocol that mitigates the issue. We observe that both the current protocol and Luo et al.'s proposal make a critical network assumption that the connections between authorities are in *bounded synchrony*. Under this assumption, any message sent by one authority reaches the other authority in a known fixed network latency (150 seconds in both cases). However, we demonstrate that this synchrony assumption is not bulletproof. An adversary, even outside of Tor's network, can carry out a well-executed denial-of-service attack to incapacitate a critical number of directory authorities, rendering the whole service unavailable.

Indeed, we show that a distributed denial-of-service (DDoS) attack can disrupt the Tor directory protocol by targeting a majority of the authorities for as little as five minutes. We conducted a controlled attack in the test environment that successfully broke the protocol. This attack is not only feasible but also cost-effective, as we estimate that it costs as little as $53.28 per month by employing standard services.

Such attacks have a profound impact on the Tor network, as its clients are instructed not to use any consensus document three hours beyond its generation to mitigate attacks using outdated consensus documents [36]. As a result, several failed consensus generation renders the whole network unavailable, as witnessed in previous outages [11, 31].

In response, this paper seeks to mitigate the problem by pivoting from the bounded synchrony network model to the *partial synchrony* network model. This model allows the network to lose synchrony temporarily due to DDoS, but eventually outputs a consensus document when the connection is restored. Drawing a parallel to the previous research, we propose a new system model dubbed *interactive consistency under partial synchrony*, and design a protocol that leverages an existing consensus protocol to solve the problem.

We implement our protocol in Rust and evaluate its performance against the current Tor directory protocol. Our results show that our protocol succeeds where the current protocol fails due to bandwidth as low as 10 Mbit/s. Our protocol introduces acceptable latency overhead compared to the current protocol, serving as a stepping stone for future work to secure the Tor directory protocol.

**Contribution.** As our first contribution, we propose a DDoS attack that paralyzes the Tor directory protocol with an extremely short time frame and only targets five servers. We successfully conducted an attack in a controlled environment and estimated the cost of the attack to be as low as $53.28 per month, using standard stressor services (see Section 4).

As our second contribution, we propose a new functionality, *interactive consistency under partial synchrony*, that models the Tor directory protocol under partial synchrony. We design a protocol that leverages existing consensus protocols to implement this functionality and give a formal security proof on its properties (see Section 5).

As our third contribution, we implement our protocol in Rust and evaluate its performance against the current Tor directory protocol as well as the previously proposed improved prototype by Luo et al. [23]. We demonstrate that our protocol continues to function under low bandwidth scenarios where both other protocols fail, while also delivering a comparable performance (see Section 6).

**Responsible Disclosure.** The current Tor directory protocol is known to be insecure as early as 2022 [23]. We disclosed our findings in this paper to the Tor project on December 30, 2024, and suggested additional anti-DDoS measures. We have not received a direct response since. Based on the new Tor DoS threat model published around the same time[1], it is stated that the remaining DoS vulnerabilities are to be addressed in the new Rust-based implementation, Arti. However, to date, Arti does not have a directory authority implementation in either its code or its roadmap. Given that Tor directory protocol is known to be vulnerable [23] and there has been little effort to discuss a true fix which needs a re-design, we hope this paper can help escalate the impact and bring the community attention to the matter. We plan to continue engaging with the Tor community as the development of Arti goes on.

## 2 Overview

The Tor project offers a volunteer-based anonymity solution where any user may direct its traffic through three intermediate proxy relays before its destination. Such proxying allows the user to effectively conceal their IP source and other metadata from the destination server. Moreover, it also allows the user to conceal the metadata of the full traffic to the intermediate relays: the first hop sees the user's IP but not the destination, and vice versa.

Nevertheless, such anonymity heavily relies on the fact that the three relays operate honestly: many attacks exist to compromise user anonymity to different degrees when one or more relays are trying to track the user [3, 34, 45]. Taken to the extreme, the user anonymity is completely lost if the three relays are controlled by the same party.

Indeed, Sybil attacks, where the attacker tries to inject as many nodes as possible into the Tor system, are a well-known threat to the Tor network [24, 44]. To compensate for this attack vector, Tor relies on having a diverse set of over 8,000 relays [40], lowering the chance for a compromised relay.

Naturally, with such a large set of relays, maintaining and ensuring its authenticity is not a trivial task: any attacker that tricks the user into using a specific set of relays can also break the user's anonymity. Therefore, Tor employs a distributed system, the *Tor Directory Protocol*, to maintain the list of relays and their authenticity.

The Tor Directory Protocol is run by a selected set of nine nodes, known as the *Directory Authorities*. As these authorities represent a critical trust and performance bottleneck in the Tor network, the directory protocol is designed to function as long as a majority of the authorities are correct [36].

[1]https://community.torproject.org/threat-model/mitigation/dos-threats/

Previously, Luo et al. [23] have demonstrated that the directory protocol is vulnerable to an equivocation attack, and suggested a fix to the protocol. However, we observe that both the current protocol and Luo et al.'s fix rely on the *bounded* synchrony assumption. In this assumption, any message sent by one authority is assumed to reach the other authorities within a fixed time-bound. The current system setting allows 150 seconds for the delivery [23]. As we will show in the next section, a well-executed attack can easily break this assumption, resulting in a complete halt of the Tor network.

## 2.1 DDoS Attack: Breaking the Synchrony

Intuitively, if an attacker concentrates its attack during the 150 seconds and launches a DDoS attack on the authorities, then the bounded synchrony assumption can be easily broken. On the other hand, every consensus protocol relies on at least a majority of the nodes being correct and responsive. Therefore, if a short DDoS attack knocks half of the authorities offline even for one timeout at the start of the protocol, then the protocol will not reach a consensus as the network assumption is broken. This result has huge implications for the Tor network, as every consensus document is set to function no longer than three hours, so consecutive failures of consensus generation will lead to a complete halt of the system. Indeed, one such attack instance has been observed in January 2021 [11]. In this instance, attackers, using different IP addresses, requested the consensus document repeatedly from the directory authority. This attack spiked the directory authority bandwidth usage to more than 400 Mbit/s, and prevented the authority from generating the consensus document for several hours.

We observe that such attacks will not be particularly difficult to execute, as the attacker only needs to target five out of the nine volunteer-based authorities, which do not have particularly robust bandwidth either (2̃50 Mbit/s as recorded [11]). Moreover, given that the authorities vote during the first 300 seconds in both the original protocol [36] and the fix [23], the attacker only needs to maintain the attack for 300 seconds on 5 authorities to completely bring down the protocol, which gives roughly $5 \times 300 = 1500$ seconds that need to be covered by the attack.

To illustrate the point, we set up a test network with nine authorities using Shadow [19], and launched a DDoS attack. We observe that the attack halts the consensus generation process, as shown in the log of one authority in Figure 1.

Furthermore, we estimate the cost of the attack using typical service costs found in previous attacks of Tor [22]. With these numbers, we observe that we can bring the network down with cost as little as $53.28 per month (See Section 4).

## 2.2 Mitigations

Given the current trust model of Tor, we cannot defend against the scenario where the attacker brings down the majority of the authorities for an extended period. Therefore, the mitigation increases the attacker's cost. Informally, we design a protocol that tolerates connectivity issues and allows the generation of a consensus document as soon as the connectivity is restored for a majority of the authorities.

```
Jan 01 01:24:30.011 [notice] Time to fetch any votes that we're
↪  missing.
Jan 01 01:24:30.011 [notice] We're missing votes from 5 authorities
↪  (1FCBF39AF0E0DB769D8124F930042FEFC6FF9FAB
↪  4075CD2D76A41DCBABF460691EBAF200FBC228F6
↪  18FB8A2D975A8B51553FA6F24845218F0D851BF7
↪  8BEA6A78258C79E7100AE0C40BF14F656BA7718D
↪  BD5D43630249C7D14FD46C8C31B1B47A442C9D02). Asking every other
↪  authority for a copy.
...
Jan 01 01:24:30.136 [info] connection_dir_client_request_failed():
↪  Giving up downloading votes from 100.0.0.8:8080
Jan 01 01:24:30.225 [info] connection_dir_client_request_failed():
↪  Giving up downloading votes from 100.0.0.6:8080
Jan 01 01:24:30.225 [info] connection_dir_client_request_failed():
↪  Giving up downloading votes from 100.0.0.7:8080
...
Jan 01 01:24:40.011 [notice] Time to compute a consensus.
Jan 01 01:24:40.011 [warn] We don't have enough votes to generate a
↪  consensus: 4 of 5
```

**Figure 1.** An example of the authority's log when five other authorities are under attack. The authority misses votes from five authorities, cannot fetch them from other authorities it can connect to, and fails to generate a consensus document based on the four votes it has received.

Simply increasing the timeout is not an effective solution. On one hand, the attacker can simply increase the attack time to match the timeout. On the other hand, since the lists of relays the authorities have are locked in at the start of the protocol, increasing the timeout also increases the latency for new relay information to be propagated to the network.

Naively, we can repeatedly run and fail iterations of some protocol until a consensus document is successfully generated. However, this approach carries a safety problem: different authorities may conclude that they have successfully generated a consensus document in different iterations, resulting in different consensus documents. This may result in an equivocation attack, as stated by Luo et al. [23], which compromises the security of the Tor network.

Fortunately, prior works have studied ways to guarantee safety even when we need to try to generate a consensus document in multiple iterations. Formally, the study on the network model of *partial synchrony* [14] allows the network to lose synchrony for a while, but eventually returns to bounded synchrony after some global synchronization time (GST) unknown to the protocol. We give a discussion of various network models in Section 3.2. It is worth mentioning that moving to the partial synchrony model will drop the fault tolerance level. Namely, in a protocol of $n$ nodes with $f$ faulty nodes, the bounded synchronous model has a theoretical bound of $f < n/2$, but the partial synchronous model can only achieve $f < n/3$ [6]. For the directory protocol of The relay is included in the consensus document if it appears at least $t \geq \lfloor n/2 \rfloor$ votes. If the relay is included, its name is determined by the vote with the largest authority ID. Its properties are determined by the popular vote, with the tie broken by the following rules.

- Each flag is not set in case of a tie.
- The largest version and/or protocol is selected.
- The lexicographically larger exit policy summary is selected.

Additionally, the relay's bandwidth is set to the median of all votes that measure them.

**Figure 2.** The process of determining the relay's properties in the Tor directory protocol [23, 36].

9 authorities, this means that we can tolerate up to 2 faulty authorities instead of 4.

Many protocols, such as PBFT [6], Tendermint [4], and HotStuff [47], have been designed to solve the consensus problem in the partial synchrony model. These protocols commonly adapt the idea of *views*, where each view represents an iteration of the protocol. The protocol eventually terminates on some view, where every authority receives the same consensus document from that view.

With these tools, we still need to characterize the Tor directory protocol and find a way to fit it into the consensus problem. Naturally, the input of the protocol is the list of relays and their properties from each authority, and the output is the consensus document, a combined list of relays and their properties. Each relay is computed separately. Based on the description of the directory protocol [36] and Luo et al.'s summarization, we give an overview in Figure 2.

Given the complexity of the aggregation, both the current protocol [36] and Luo et al.'s fix [23] opted to characterize the protocol as a broadcast problem, where each authority broadcasts its state to the other authorities in parallel. After that, each authority aggregates the received states locally, signs on the aggregated result, and broadcasts the signature. Given that the states are broadcast to all authorities, the protocol ensures that each authority has the same set of states, and the aggregated result is the same for all authorities.

Unfortunately, we observe that the Byzantine broadcast problem under partial synchrony can only guarantee delivery of the broadcast message if GST is zero, or, in other words, the network never lost synchrony [1]. The reason is that otherwise, the receiver cannot be sure if the correct sender has sent the message but is stuck in some unknown GST (in which case it should wait) or if a faulty sender has not sent the message at all (in which case it should proceed). This is undesirable for the Tor directory protocol, which needs to generate a non-empty consensus document even if the network has lost synchrony for a while.

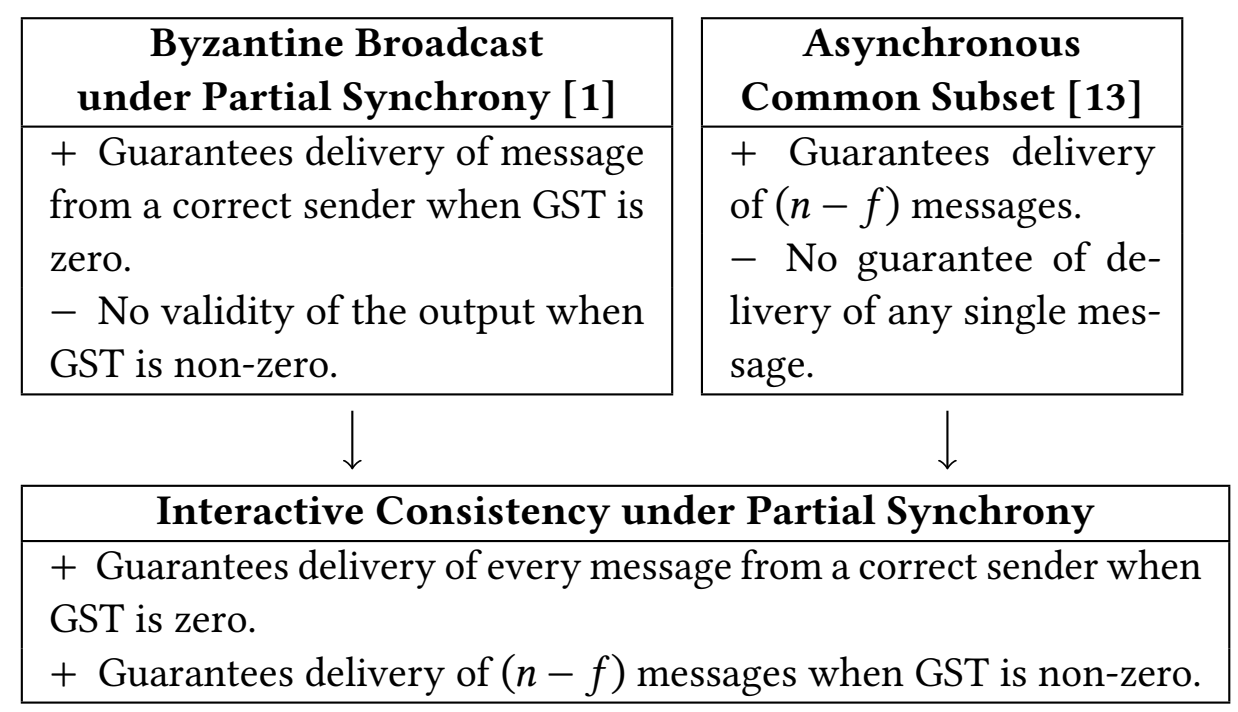

**Figure 3.** The functionality we proposed can be considered as a combination of Byzantine Broadcast under Partial Synchrony and Asynchronous Common Subset. Interactive Consistency under Partial Synchrony guarantees delivery of every message when GST is zero, and delivery of $(n - f)$ messages when GST is non-zero.

To deal with this problem, we borrow the idea from the asynchronous common subset [13]. Specifically, while any one specific broadcast is not guaranteed to succeed, we can still guarantee that $(n - f)$ broadcasts will succeed before we move on. Therefore, each authority will still receive the input from $n - f \geq 2f + 1$ authorities, in which only up to $f$ are faulty. This guarantees that correct inputs will outweigh faulty ones, ensuring that both majority votes and median values will output the values based on correct inputs.

Putting these parts together gives us the best of both worlds. On one hand, if the protocol is not under attack and the network is synchronous, we can leverage broadcast properties and ensure that the input from every correct authority propagates to the output. On the other hand, if the network loses synchrony for any reason, such as a DDoS attack, we can still guarantee that the protocol still produces an output that consists of a majority of correct inputs.

Following the definition of Byzantine broadcast under partial synchrony [13], we formalize this functionality as *Interactive Consistency under Partial Synchrony* (Definition 5.1). Informally speaking, in this functionality, each node will receive the message from every correct node if the network is synchronous (known as interactive consistency [10]). On the other hand, when GST is non-zero, the nodes are still guaranteed to receive messages from $(n - f)$ nodes. We give an overview of the functionality in Figure 3. This definition is general and applicable to other similar settings.

We then show that we can leverage existing consensus protocols using views to solve this problem by adding a dissemination phase and an aggregation phase (see Section 5).

We implemented our protocol in Rust and evaluated its performance. Our result shows that our protocol performs at a similar level to the current Tor directory protocol, introducing an additional overhead of seconds in the context of the original 10-minute protocol. We observe that our protocol separates the process of sending relay lists and achieving consensus, allowing for an arbitrary timeout while sending the file and only requiring partial synchrony when reaching the consensus. As a result, our protocol tolerates low-bandwidth scenarios where the current protocol would fail due to its synchrony assumption being violated (see Section 6).

# 3 Preliminary

We start by giving an overview of the current Tor directory protocol and state its system model, and then discuss the network model and the security definition we will be using in this paper.

## 3.1 Tor Directory Protocol

Since the functionality of Tor relies on volunteer-operated relays, it's crucial for the user to have an updated list of these relays for circuit establishment. Ensuring users have accurate relay information is vital for anonymity, as collusion between different relays compromises user identity.

Traditional solutions fall short. Embedding the list of relays in the software isn't feasible given their changing status, and pulling this information from a central source like Tor's website creates a vulnerability. If the site goes down or is compromised, the entire system is at risk.

To mitigate these issues, Tor uses a decentralized network of nine specialized directory authorities that collect and distribute relay-node data. Although a client could theoretically gather this data from each authority individually, it's more efficient to retrieve one consolidated document due to bandwidth constraints, particularly given Tor's large user base. This led to the creation of a directory protocol that allows authorities to reach a consensus on relay information and create a so-called consensus document every hour.

The directory protocol has undergone several iterations and is currently in its version 3. This version comes with an exhaustive set of guidelines detailing how authorities should behave [36]. Additionally, as an open-source initiative, Tor makes its code, including the directory authority component, publicly available for scrutiny [37]. We outline the protocol's four phases in Figure 4.

Given that users using different consensus documents exhibit different traffic patterns, Tor enforces the use of the latest consensus document rigorously [36]. Tor treats a consensus document as stale one hour after its generation, advising clients to avoid using it. However, the document remains valid for up to three hours. Beyond this three-hour threshold, the consensus document is deemed invalid. This rule renders the proper and timely generation of consensus documents extremely important, as a sustained lack of consensus documents for as little as three hours renders the whole network invalid, as demonstrated in one outage [11].

| Step 1. Perform Vote | Step 2. Fetch Votes |
| --- | --- |
| Each authority sends its list of relays as a vote to every other authority. | If an authority has not received a certain vote, it tries to fetch the vote from every other authority. |
| **Step 3. Send Signature** | **Step 4. Fetch Signatures** |
| Each authority aggregates the votes locally, signs the resulting consensus document, and sends the signature to every other authority. | If an authority has not received a certain signature, it tries to fetch the signature from every other authority. |

**Figure 4.** The current Tor directory protocol consists of four rounds, with each round lasting 150 seconds. The protocol is executed once every hour.

Version 3 was considered secure until recently when Luo et al. [23] exposed a potential vulnerability. They demonstrated an equivocation attack that allows a compromised authority to tamper with the consensus document, thereby stripping users of their anonymity. As a remedy, the work proposed an improved protocol that leverages a Dolev-Strong style consensus [12]. Figure 5 demonstrates an outline.

We note that both the original protocol and Luo et al.'s proposed protocol assume a bounded synchrony of 150 seconds, a parameter we aim to exploit in this paper. We also note that both protocols transfer the relay information during the first two rounds. Therefore, we can focus our DDoS attacks on these two rounds to bring the protocol down.

As of 2024, the Tor network operates with nine directory authorities [40]. The migration of Tor's codebase from C to Rust is in progress, and thus, it is anticipated that version 3 of the directory protocol will continue to be in use for the foreseeable future.

## 3.2 Bounded Synchrony, Asynchrony and Partial Synchrony

*Bounded synchrony* [12] is a foundational assumption prevalent in network models that guide various protocols. This model stipulates that a message sent by a correctly functioning node will reach another such node within a predetermined latency window $\Delta$. The current Tor directory protocol has relied on this assumption, with the currently deployed parameter of 150 s.

On the other hand, *asynchronous* models make no assumptions about the network's timeout and allow messages to be arbitrarily delayed before they are delivered. While this model is the most general one, the famous FLP impossibility [16] dictates that no deterministic consensus protocol can be designed in an asynchronous network. As a result, consensus protocols under asynchronous models have to rely on randomness, most commonly under the form of a common coin [8, 9] [2]. Such designs significantly increase the complexity of the protocol and lower the performance, and so we elect not to adopt them.

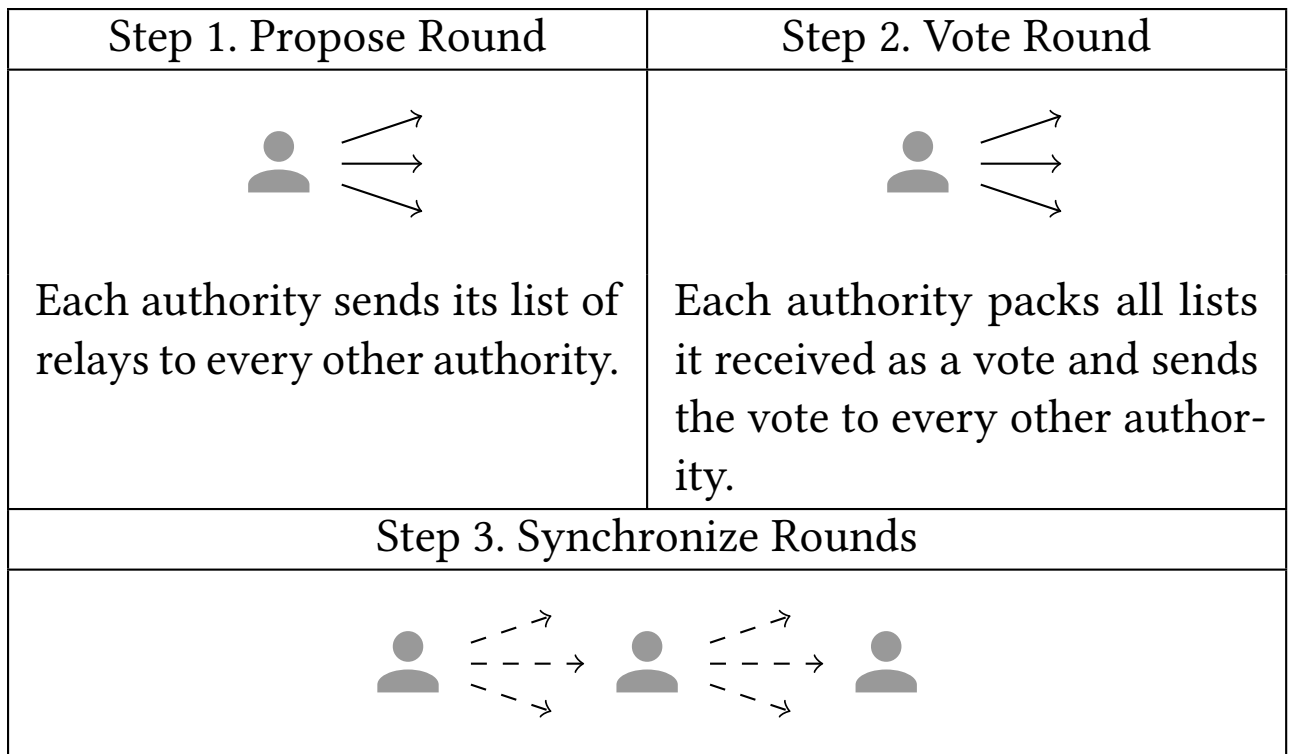

**Figure 5.** The improved Tor directory protocol proposed by Luo et al. [23]. This protocol also assumes a bounded synchrony of 150 seconds.

In contrast, the concept of *partial synchrony* [4, 6, 47] offers a more flexible model to account for network variability. The central tenet of partial synchrony is that a protocol should maintain its security even when the network is under adversarial control and should regain liveness once ideal network conditions are restored.

Formally, partial synchrony introduces the notion of a Global Stabilization Time (GST), during which communication between nodes is not guaranteed and may occur in an arbitrary order. The protocol does not know the duration of GST in advance, making it unable to adjust its behavior based on this period. Essentially, GST serves to model scenarios where the network experiences fluctuations or is under adversarial influence. Following the conclusion of GST, the network reverts to a more predictable state, characterized by a known latency $\Delta$ between nodes.

### 3.3 Consensus Problem and View-based Protocols

Informally speaking, the consensus problem refers to the challenge of achieving agreement among a group of nodes on a single value, given that a minority of the nodes are faulty and may behave arbitrarily. One formalization of the problem is the Byzantine agreement problem, which we present a definition below [30].

**Definition 3.1** (Byzantine Agreement)**.** *There are $n$ nodes, each with an input value. Up to $f$ of them can be faulty. A solution needs to satisfy the following requirements:*

*(1)* ***Termination.*** *All correct nodes eventually output a value.*
*(2)* ***Agreement.*** *All correct nodes output the same value.*
*(3)* ***Validity.*** *If all correct nodes hold the same input value $v$, then they all output $v$.*

Under the partial synchrony model, the theoretical tolerance threshold for Byzantine agreement is $n \geq 3f + 1$. Many protocols use the concept of *views* to achieve this bound, such as PBFT, Tendermint, and HotStuff [4, 9, 47].

From a high level, a view is a period of time during which a predetermined leader tries to achieve agreement among the nodes. If the leader is faulty or if the network times out, a *view-change protocol* is initiated, where nodes communicate with the leader of the next view to ensure that they have enough information to not violate the agreement. Then the protocol moves on to the next view. This process repeats until the nodes reach agreement.

The exact method of reaching agreement in each view and the view-change protocol vary between different protocols. PBFT is the first to motivate the concept, but uses quadratic communication per view [9]. Tendermint reduces the communication complexity to linear but requires the protocol to wait during the view-change protocol to ensure that the leader of the next view has enough information [4]. HotStuff addresses the issue to avoid any waiting time but has a higher round complexity [47].

### 3.4 Variants of Consensus Problems and Interactive Consistency

A few other variants of the consensus problem motivate the functionality we define in this paper. The (bounded synchronous) Byzantine broadcast problem characterizes the scenario where a potentially faulty sender sends a value to all other nodes, and the nodes need to agree on the value.

**Definition 3.2** (Byzantine Broadcast [1])**.** *There are $n$ nodes and a designated sender among them in a bounded synchronous network. Up to $f$ of them can be faulty. A solution needs to satisfy the following requirements:*

*(1)* ***Termination.*** *All correct nodes eventually output a value.*
*(2)* ***Agreement.*** *All correct nodes output the same value.*
*(3)* ***Validity.*** *If the designated sender is honest, then all correct nodes output the sender's value.*

Unfortunately, under the partial synchronous network model, message delivery can only be guaranteed when GST is zero [1]. A formal definition of the problem under partial synchrony is given below.

**Definition 3.3** (BB under Partial Synchrony [1])**.** *There are $n$ nodes and a designated sender among them in a partial*

[2] While Tor has a common coin for its hidden services, the coin is bootstrapped from the consensus document, so having the consensus document rely on this common coin creates a cyclic dependency.

*synchronous network. Up to $f$ of them can be faulty. A solution needs to satisfy the following requirements:*

(1) ***Termination.*** *All correct nodes eventually output a value.*
(2) ***Agreement.*** *All correct nodes output the same value.*
(3) ***Validity.*** *If the designated sender is honest and GST is zero, then all correct nodes output the sender's value.*

On the other hand, the asynchronous common subset (ACS) problem is a variant of the consensus problem that requires the nodes to agree on a set of values. We give a definition below. For this definition, we denote $|V|_{\neq\perp}$ as the number of non-empty elements in vector $V$.

**Definition 3.4** (Asynchronous Common Subset [13])**.** *The protocol runs with a system of $n$ nodes $\{P_1, P_2, \ldots, P_n\}$ with up to $f$ being faulty under an asynchronous network. Each node $P_i$ starts with a value $x_i$. The following properties hold after the protocol execution:*

(1) ***Termination.*** *Every correct node $P_i$ outputs a vector $X_i = \{x_{i,1}, x_{i,2}, \ldots, x_{i,n}\}$ of size $n$, where $x_{i,j}$ is either a value or $\perp$.*
(2) ***Agreement.*** *If node $P_i$ and $P_j$ are correct, then $X_i = X_j$.*
(3) ***Validity.*** *If node $P_i$ is correct, then $|X_i|_{\neq\perp} \geq n - f$.*

Finally, interactive consistency, the problem that Luo et al.'s work on the Tor directory protocol [23] aims to solve, is a variant of the consensus problem under bounded synchrony that allows every node to broadcast a value, and decide on a set that consists of these values. We give a definition below.

**Definition 3.5** (Interactive Consistency [15])**.** *The protocol runs with a system of $n$ nodes $\{P_1, P_2, \ldots, P_n\}$ with up to $f$ being faulty under a bounded synchronous network. Each node $P_i$ starts with a value $x_i$. The following properties hold after the protocol execution:*

(1) ***Termination.*** *Every correct node $P_i$ outputs a vector $X_i = \{x_{i,1}, x_{i,2}, \ldots, x_{i,n}\}$ of size $n$, where $x_{i,j}$ is either a value or $\perp$.*
(2) ***Agreement.*** *If node $P_i$ and $P_j$ are correct, then $X_i = X_j$.*
(3) ***Validity.*** *If a correct node $P_i$ starts with the value $x_i$, then $x_{i,i} = x_i$.*

As we will see later, our definition of interactive consistency under partial synchrony (Definition 5.1) combines the ideas of these problems to derive a definition of the same problem under partial synchrony.

## 4 Exploiting the Synchrony Assumption

To demonstrate the weakness of the synchrony model currently in use, we will show that the attacker can exploit the model and employ distributed denial-of-service (DDoS) attacks to render the whole Tor system unresponsive by flooding the directory authorities.

### 4.1 Threat Model and Assumptions

In this attack, we do not require the attacker to be or control any directory authority, like prior attacks [23]. In contrast, the attacker can be an outsider of the Tor network.

Similar to prior works [22], we assume the attacker employs DDoS-for-hire stressor services to launch DDoS attacks against the directory authorities whose IPs are public. The stressor services use a distributed botnet of compromised hosts to flood the target with requests. Following the DoS attacking model proposed by Jansen et al. [22] during its analysis of the Tor system, a target will have reduced available bandwidth for performing the directory protocol if it was suffering from the DDoS attack.

### 4.2 Attack's Idea

We recall from Section 3.1 that the current directory protocol is operated by 9 directory authorities and operates in four 150-second rounds to generate one consensus document every hour: Perform Vote, Fetch Votes, Send Signature, and Fetch Signatures. As long as authorities generate a consensus document and receive at least 5 signatures (the majority of 9 authorities) on it, the consensus document is valid, and the consensus process is considered successful.

We observe that the protocol uses the first two rounds to send vote files. Therefore, the protocol will fail if the attacker stops a majority of the authorities from sending messages in the first two-round interval (300 seconds) successfully. To achieve this goal, the attacker can launch a DDoS attack against the majority (5 out of 9) of the authorities during the time. Authorities under the DDoS attack only have limited bandwidth for message delivery, and therefore cannot collect 5 votes to generate a valid consensus document within the bounded time. As a result, the directory protocol will fail.

### 4.3 Attack Evaluation and Cost

Given that it is difficult to assess DDoS attacks in real life since carrying out the attacks on any target creates stress on ISP and thus ethic concerns, existing works [22] focus on a simulated environment and show that attacking a small number of targets has a disproportionate effect on the big system. Following this method, to evaluate the impacts of DDoS attacks on the current Tor protocol, we use Shadow [19] to conduct our experiments. Shadow is a high-fidelity network simulator that enables us to run the Tor system with customized network topology and link bandwidth between nodes, and has been adopted to analyze the DoS attacks against Tor by previous work [22]. For the following evaluation, we use a server to run the Tor directory protocol on Shadow. The server runs Ubuntu 22.04 and is equipped with 48 CPU cores, 128 GB of RAM, and 10 TB of SSD.

**Network Setup.** We use the Tor network generation tools [20] to create a private Tor network with realistic history data of the live Tor network. The private network simulates the realistic network topology and traffic between Tor nodes [22]. We then use the private network to perform the directory protocol on Shadow. Similar to experiment settings in [22], instead of directly launching the DDoS attack, we simulate the attack on authorities by limiting their available bandwidth. In the following evaluation, attacked authorities will be using their limited bandwidth to perform the directory protocol, including disseminating their vote files and servicing for vote requests.

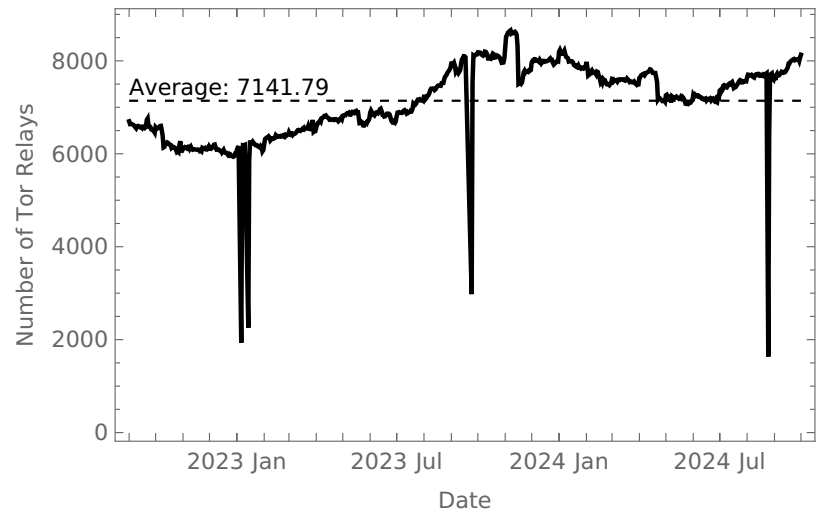

**Figure 6.** The number of Tor relays over time since September 2022 [40]. The dashed line represents the average number of relays (7141.79).

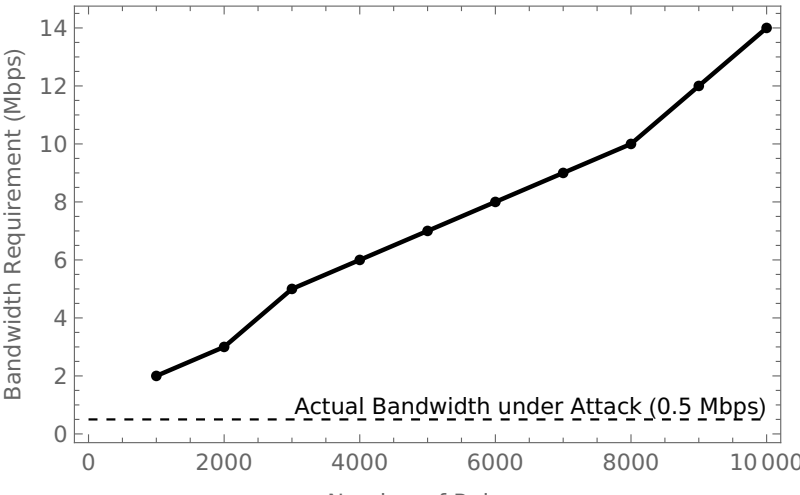

**Figure 7.** The bandwidth requirements for authorities to perform the directory protocol successfully under distinct numbers of relays. Dashed line represents the available bandwidth for a node under DDoS attack [22].

As the directory protocol requires authorities to share their vote files, we also need to measure the size of the vote files. Specifically, a voting document created by an authority contains some meta-data (voting interval, length of each step, etc.) and information on every relay the authority knows. As a consequence, the size of a shared vote file is determined by the number of live relays that an authority interacts with. To this end, we collect the historic numbers of live relays in the Tor network from Tor Metric [40]. Figure 6 illustrates the number of relays on the Tor network over the last two years (from September 2022 to October 2024). We find that the number of relays is changing over time, with an average of around 7,100. This dynamic of relays motivates us to evaluate distinct bandwidth requirements for authorities to perform the directory protocol successfully, given a varying number of relays in the system.

**Evaluating Bandwidth Requirements.** Figure 7 presents the authorities' bandwidth requirements for successfully performing the directory protocol under distinct numbers of relays. Specifically, we use different numbers of relays (denoted by $r$, from 1,000 to 10,000) to generate varying sizes of a vote file for each authority. We then run 9 authorities on Shadow to perform the directory protocol, where 5 of the 9 authorities are set with limited bandwidth as they are suffering from bandwidth DoS attacks. Therefore, the required bandwidth represents the (almost) minimum bandwidth required by each of the 5 authorities such that the directory protocol completes successfully. From Figure 7, we conclude that the required bandwidth of authorities is proportional to the number of relays. To elaborate on the evaluation results, from the attacker's perspective, an attack is guaranteed to succeed if it limits the available bandwidth of 5 authorities to be lower than the given required bandwidth. For instance, given the current 8,000 relays in the network, each authority requires at least around 10 Mbit/s (1.25 MB/s) during the period when it participates in the first two rounds of the directory protocol. Previous works estimate that 0.5 Mbit/s will be available to a node during a DDoS attack [22], far below the necessary requirement for the directory protocol. Therefore, we conclude that our attack is effective in disrupting the directory protocol.

**Attack Cost.** We now estimate the attack cost. Specifically, we need to estimate the link capacities of authorities and the costs required to subscribe to DDoS-for-hire stressor services. Tor authorities do not explicitly publish their link capacities. To estimate a given authority's link capacity, we refer to the post [11] that reports a 250 Mbit/s bandwidth setting for the authority's link. Observation from bandwidth measurements done by bandwidth authorities suggests around the same bandwidth [35]. Given that an authority requires 10 Mbit/s bandwidth to perform the synchronous directory protocol including 8,000 relays, an attacker can flood a single authority with 240 Mbit/s of attack traffic such that the authority has insufficient bandwidth for the directory protocol. For denial-of-service attack costs, we use data from Jansen et al. [22]. More specifically, the average amortized cost of the attacker to flood a single target with 1 Mbit/s of attack traffic for an hour is \$0.00074. Applying this cost model, we estimate the attack costs that attackers use stressor services to flood 5 authorities for 5 minutes to breach one instance of the directory protocol is approximately \$0.074. Since authorities initialize a consensus instance of the synchronous directory protocol per hour, the estimated cost to crash the Tor network by breaching all consensus runs would be approximately \$53.28/month.

## 5 New Protocol for Tor Directory

We first discuss the system model and requirements for the new protocol, and then give a construction that satisfies the requirements. Finally, we analyze the security and performance of the protocol.

### 5.1 System Model and Requirements

**Input and Output.** To design a secure protocol for the Tor directory protocol, we must consider the design goals of the system. The input of the protocol is the status document of each authority, depicting the list of relays the authority knows. The output of the protocol is the consensus document, a combined list of relays and their properties.

Unfortunately, Tor's algorithm for aggregating the status document is quite complicated, involving the evaluation of many parameters based on the properties of the relays [36]. As a result, all existing implementations [23, 37] decide to treat the protocol as a broadcast problem of status votes and have authorities aggregate these votes locally via a deterministic algorithm. A full treatment of the design of such an algorithm needs to balance many aspects of Tor going beyond the scope of the paper. Nevertheless, given that the directory protocol is designed to safeguard against a minority of authorities being faulty, for this paper, we assume this aggregation algorithm is robust: as long as the input contains more votes from correct authorities than from faulty ones, the output will make sense.

Therefore, the input and output for our problem is similar to that of the interactive consistency problem (Definition 3.5), where each authority starts with a value (the status document) and the output is a vector of values.

**Problem Definition.** Based on the above discussion, we now provide a formal definition of the problem we are trying to solve. While prior work [23] model the problem as an interactive consistency problem under bounded synchrony, the exact same definition is not possible under partial synchrony, since we cannot ensure that any broadcast message is guaranteed to be delivered [1]. Nevertheless, by combining the idea from interactive consistency and Byzantine broadcast under partial synchrony, we propose *interactive consistency under partial synchrony* that gives us the best of both worlds. We ensure the delivery of every message when the network is synchronous, and that the protocol still produces an output that consists of at least $(n-f)$ inputs when the network loses synchrony, which ensures that the aggregated result will be sound. We present this formal definition below. For this definition, we denote $|V|_{\neq\perp}$ as the number of non-empty elements in vector $V$.

**Definition 5.1** (Interactive Consistency under Partial Synchrony)**.** *The protocol runs with a system of n nodes* $\{P_1, \dots, P_n\}$ *under partial synchrony and a Byzantine fault tolerance of f. Each node* $P_i$ *starts with a value* $x_i$*. The following properties hold after the protocol execution:*

*(1)* ***Termination.*** *Every correct node* $P_i$ *outputs a vector* $X_i = \{x_{i,1}, x_{i,2}, \dots, x_{i,n}\}$ *of size n, where* $x_{i,j}$ *is either a value or* $\perp$*.*
*(2)* ***Agreement.*** *If node* $P_i$ *and* $P_j$ *are correct, then* $X_i = X_j$*.*
*(3)* ***Value Validity.*** *If a correct node* $P_i$ *starts with the value* $x_i$*, then* $x_{i,i} \in \{x_i, \perp\}$*. Specifically, if* $GST = 0$*, then* $x_{i,i} = x_i$*.*
*(4)* ***Common Set Validity.*** *If node* $P_i$ *is correct, then* $|X_i|_{\neq\perp} \geq n - f$*.*

**Fault Tolerance.** Formally, the current protocol deployed by Tor solves the interactive consistency problem and assumes an $n \geq 2f+1$ fault tolerance [23], which is the optimal threshold for the bounded synchrony model it operates under. As we have shown above, the bounded synchrony assumption is not realistic under DDoS attacks, which can cause the network to lose synchrony. Therefore, in this section, we explore a more relaxed network model, partial synchrony, and design a new protocol that can operate under this model. However, in order to move to a more relaxed network model, we must reduce the fault tolerance to $n \geq 3f + 1$, which is the theoretically optimal threshold for partial synchrony, to ensure the protocol's security. This dilemma presents a natural trade-off between the network assumptions and the fault tolerance threshold.

In Tor's setting, reducing fault tolerance can be preferable if it yields stronger resilience. As demonstrated above, DDoS attacks completely invalidate the bounded synchrony assumption, so we must consider a more relaxed network model to build resilience against such attacks. Furthermore, Tor is known to vet its directory authorities extensively, sometimes described as "super trusted we-know-you-and-have-had-many-beers-with-you [41]." The permissioned, high-trust structure shifts the design trade-off toward resilience against uncontrollable DDoS risk rather than tolerance of multiple adversarial authorities. It is also acknowledged in the directory authority expectation that "a single bad directory authority shouldn't be able to do much harm, but a few acting together can [36]." The wording suggests that tolerating a significant number, say four out of nine, adversarial directory authorities is not the main design goal of the directory protocol.

We will present a protocol below that leverages an established consensus protocol under partial synchrony.

### 5.2 Protocol Description

In our protocol, we divide the execution into three sub-protocols: dissemination, agreement, and aggregation.

The dissemination sub-protocol processes each node's input document to generate a digest vector alongside an externally verifiable proof that the digest is correct $(H, \pi)$ for the leader. This pair $(H, \pi)$ is then channeled into the agreement sub-protocol as the input for the leader of the consensus.

The agreement sub-protocol takes the pair $(H, \pi)$ as the input, and aims to get every node to agree on this pair. This is equivalent to solving the consensus problem under partial synchrony, and we can utilize any view-based consensus protocol, such as PBFT [6], Tendermint [4], or HotStuff [47].

Upon concluding the agreement sub-protocol, every correct node agrees on a digest vector $H$. The subsequent step involves recovering the document from the vector and securing every node's signature on the aggregated consensus document, managed within the aggregation sub-protocol. An outline of the process is available in Figure 8.

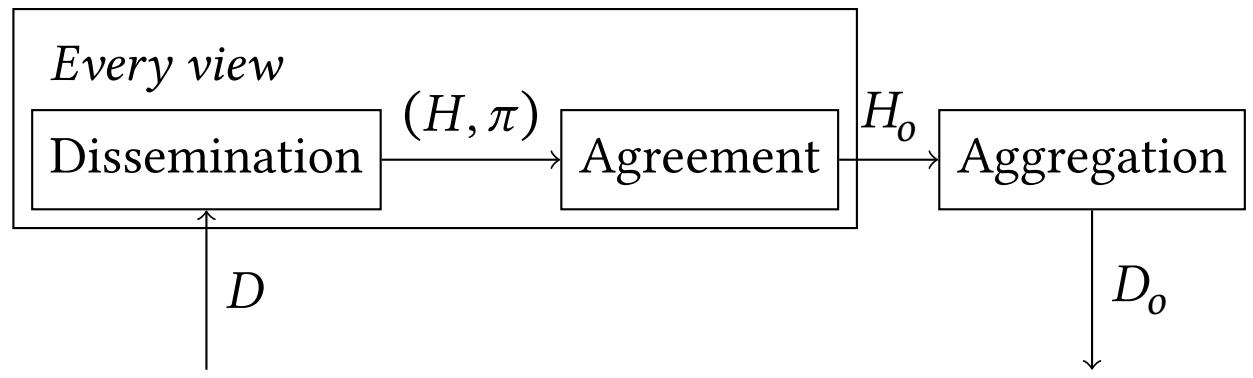

**Figure 8.** A schematic representation of the protocol. Initially, the protocol takes a document from every node, forming the input vector $D$. 1. For every view, the dissemination sub-protocol generates a digest vector $H$ and an externally verifiable proof $\pi$ for the leader; 2. Subsequently, the agreement sub-protocol takes $(H, \pi)$, yielding a consensus digest vector $H_o$ acknowledged by all correct nodes; 3. Lastly, the aggregation sub-protocol utilizes $H_o$ to reconstruct the output vector $D_o$, ensuring consistency across all correct nodes.

We give a detailed description of each sub-protocol.

**5.2.1 Dissemination.** The sub-protocol disseminates the documents and produces a digest vector $H$ and a proof $\pi$ for the view leader to be used in any consensus protocol.

At the start, every node $i$ broadcasts its document $d_i$ with the digest $h_i$ and a signature $\langle \text{DOCUMENT}, d_i, h_i, \sigma_i(i, h_i)\rangle$ to every node. Each node waits until either

1. they have received all $n$ DOCUMENT messages, or
2. the timeout $\Delta$ has passed, and they have received at least $(n - f)$ DOCUMENT messages.

At the start of every view with leader $l$:

1. For every node $i$, if it has received at least $(n - f)$ DOCUMENT messages since the start of the protocol, it creates a proposal

$$\begin{aligned} P_i = ((&h_{i,1}, \sigma_1(1, h_{i,1}), \sigma_i(1, h_{i,1})), \\ &(h_{i,2}, \sigma_2(2, h_{i,2}), \sigma_i(2, h_{i,2})), \\ &\dots, \\ &(h_{i,n}, \sigma_n(n, h_{i,n}), \sigma_i(n, h_{i,n}))) \end{aligned}$$

   with the following rules:
   a. $h_{i,j} = h_j$ if node $i$ received a valid DOCUMENT message from node $j$;
   b. $h_{i,j} = \perp$ otherwise.

   Note that a proposal $P_i$ must have at least $(n - f)$ non-empty entries to be valid since the node only creates a proposal if it received at least $(n - f)$ DOCUMENT messages. Every node $i$ then sends $\langle \text{PROPOSAL}, P_i\rangle$ to the leader $l$.
2. Upon having received a new PROPOSAL message and cumulatively at least $(n - f)$ PROPOSAL messages this view, the leader $l$ creates a vector $H_l$ and a proof $\pi_l$ with the proposals it received $P\,(|P|_{\neq\perp} \geq n - f)$ as:
   a. $H_{l,j} = h_j$ if there are at least $(f+1)$ proposals with $h_j$ as the corresponding value for $j$. In this case, the corresponding proof $\pi_{l,j}$ is the vector of the signatures $\sigma(j, h)$ from $(f + 1)$ nodes. The $(f + 1)$ limitation assures that at least one correct node has the full document, and the document may be retrieved from this node as necessary.
   b. $H_{l,j} = \perp$ if there exists an equivocation in $j$. In this case, the corresponding proof $\pi_{l,j}$ is any two different message digests $(h, h')$ and their signature by the sender $(\sigma_j(j, h), \sigma_j(j, h'))$.
   c. $H_{l,j} = \perp$ otherwise. Since there are at least $(n - f)$ proposals, if there is no equivocation, there are at least $(f + 1)$ proposals with $\perp$ as the corresponding value for $j$. Therefore, the corresponding proof $\pi_{l,j}$ is the vector of the signatures $\sigma(j, \perp)$ from $(f + 1)$ nodes. The $(f+1)$ limitation assures that at least one correct node has not received the full document, so an adversarial leader cannot exclude correct nodes when $GST = 0$.

   The leader $l$ then determines $H_l$ to be *ready* only if it has at least $(n - f)$ non-empty elements. That is, $|H_l|_{\neq\perp} \geq n - f$. Only a ready digest vector will be used for the agreement sub-protocol input. If $H_l$ is not ready, the leader waits for more PROPOSAL messages before entering the agreement sub-protocol.

Refer to Figure 9 for a visual representation.

**5.2.2 Agreement.** The agreement sub-protocol can be any view-based Byzantine Agreement protocol that works under partial-synchrony. The input of the protocol is the input vector $H$ of size $O(n)$, with a proof $\pi_i$ of size $O(n)$ for external validity. The output of the protocol is the output vector $H_o$ of size $O(n)$, an input vector agreed by every correct node.

**5.2.3 Aggregation.** At the end of the sub-protocol, every node has a digest of all the documents included in the output. They then fetch the documents corresponding to the digest from every other node if they have not received them from other nodes before. Since the dissemination sub-protocol ensures that the digest is correct, the document is guaranteed to be available from at least one correct node.

After that, the interactive consistency under partial synchrony protocol finishes. Authorities running the protocol aggregate the documents to obtain the consensus document locally, using the current Tor directory protocol's aggregation algorithm, and then sign and broadcast the signature.

### 5.3 Analysis of the Protocol

We delegate the relatively straight proof that our protocol adheres to our definition of interactive consistency under partial synchrony to Appendix A.

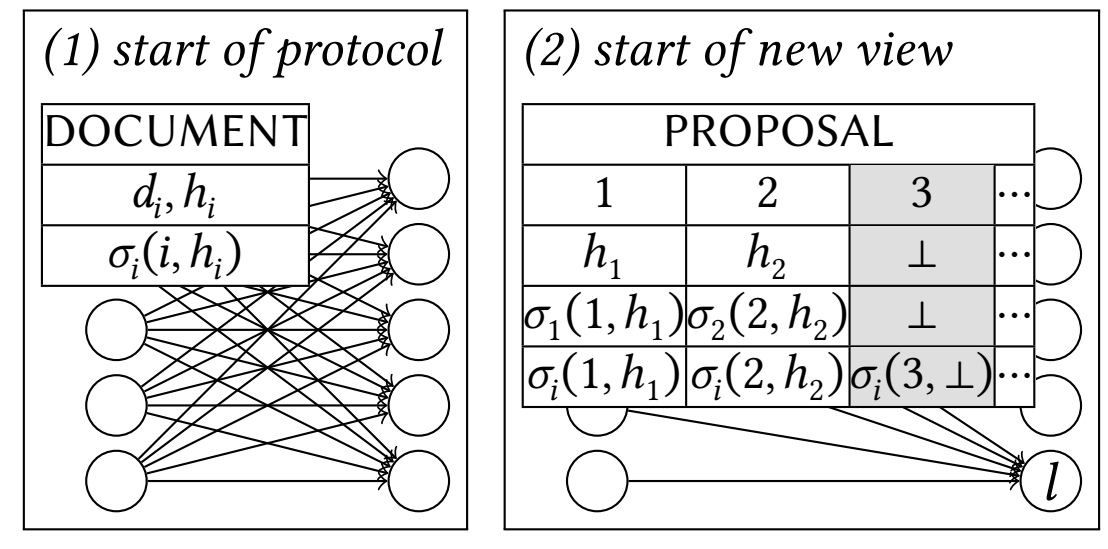

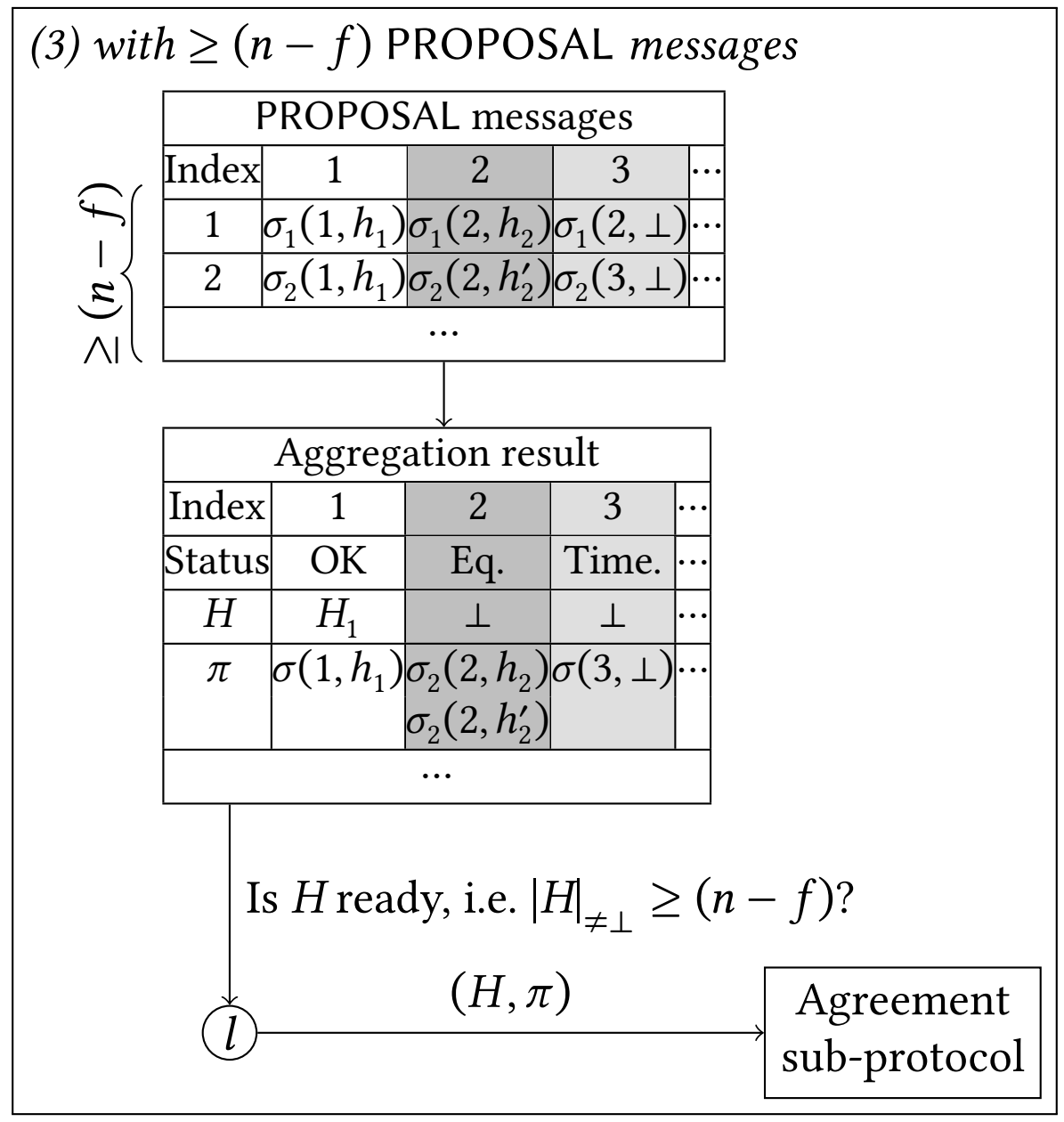

**Figure 9.** Overview of the Dissemination sub-protocol. Light columns indicate timeouts and dark columns denote detected equivocations. 1. Nodes begin by broadcasting a DOCUMENT message, and proceed when they have received enough DOCUMENT messages; 2. At the start of each new view in the agreement sub-protocol, nodes, upon receipt of at least $(n - f)$ DOCUMENT messages, send a PROPOSAL message to the leader; 3. The leader aggregates PROPOSAL messages, setting them as the agreement sub-protocol input when it is ready.

We conduct a complexity analysis of each sub-protocol in Appendix B. Denote $d$ as the document size and $\kappa$ as the signature size. We achieve $O(n^2 d + n^3 \kappa) + O(\text{Agreement})$ communication complexity. Our protocol adds 4 rounds of overhead compared to the consensus protocol.

## 6 Evaluation

We implemented our protocol in Rust based on a variant of HotStuff [17]. Overall we added around 1000 lines to the original codebase to implement our functionality. Our code is available at https://github.com/zhtluo/partialtor-rs.

### 6.1 Experiment Setup

Similar to the evaluation of the attack, we evaluate our protocol and compare it with the current Tor directory protocol [37] as well as Luo et al.'s synchronous prototype [23] in Shadow [19], a high-fidelity network simulator. To derive the real network latencies among the 9 authorities, we use Tornettools [20] to create a private Tor network for Shadow. For the current Tor directory protocol and Luo et al.'s synchronous prototype, we use the same synchronous settings adopted by the current Tor network. In this setting, the authority runs one round in 2.5 minutes and finishes the protocol with 4 rounds in 10 minutes (see Section 3.1). For our protocol, however, there is no such lock-step setting as it was designed to work under a partially synchronous network.

We run our evaluations on a server with 48 CPU cores, 128 GB of RAM, and 10 TB SSD.

### 6.2 Performance

We compare our protocol with the current Tor directory protocol and Luo et al.'s implementation in terms of latency. Latency generally refers to the time between the start of the protocol and the generation of the consensus document. However, in the current directory protocol and the synchronous prototype, a consensus document is generated in a fixed time (i.e., 4 lock-step rounds in 10 minutes), while our protocol does not employ lock-step rounds due to the design. To fairly compare the time used to generate a consensus document, we calculate the latency of such protocols by adding up the processing time of each round, effectively computing its network time. For instance, for the first vote round, we consider its processing time from when an authority posts its vote document to when the authority receives all votes.

Figure 10 shows the latency comparison between the current Tor directory protocol (denoted 'Current'), Luo et al.'s implementation (denoted 'Synchronous'), and our protocol (denoted 'Ours') under different bandwidth settings and different numbers of relays. Specifically, we set the number of relays to be between 1,000 and 10,000 and the bandwidth of the authorities to be at 50 Mbit/s, 20 Mbit/s, 10 Mbit/s, 1 Mbit/s, and 0.5 Mbit/s.

We find that the current directory protocol breaks down when the number of relays exceeds its capacity. To elaborate, authorities in the current Tor directory protocol require enough bandwidth resources to disseminate vote documents in time. Recall from our attack evaluation in Section 4.3 that an authority requires more than 10 Mbit/s bandwidth to serve 8,000 relays. Thus, the current Tor directory protocol fails to generate a consensus document when there exist between 9,000 and 10,000 relays in the network. Luo et al.'s protocol fares worse, and fails to generate a consensus document for over 1,000 relays under the same bandwidth. We think the increased complexity in their implementation (see Table 1

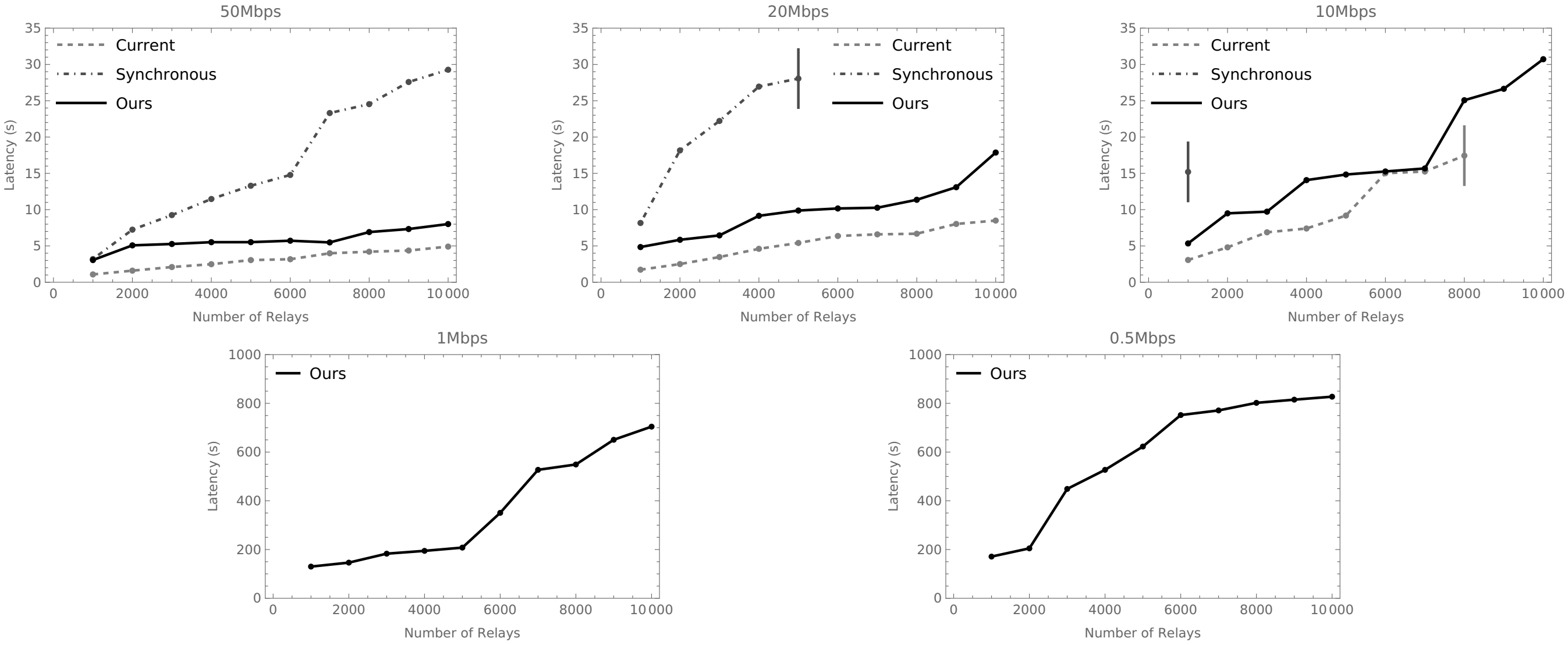

**Figure 10.** Latency of generating a consensus document between the current directory protocol [37], the synchronous directory protocol proposed by Luo et al. [23], and our protocol. We measure the network time of generating a consensus document under different bandwidth settings (50 Mbit/s, 20 Mbit/s, and 10 Mbit/s, respectively) with different numbers of relays. Both the current protocol and the synchronous protocol fail to generate a consensus document under lower bandwidth settings when the number of relays exceeds certain thresholds (denoted as thick vertical lines in the figure). Notably, the synchronous protocol fails to generate a consensus document under 10 Mbit/s bandwidth when the number of relays exceeds 2,000; both the current protocol and the synchronous protocol fail to generate a consensus document under 1 Mbit/s and 0.5 Mbit/s bandwidth with 1,000 relays, the lowest number of relays we test.

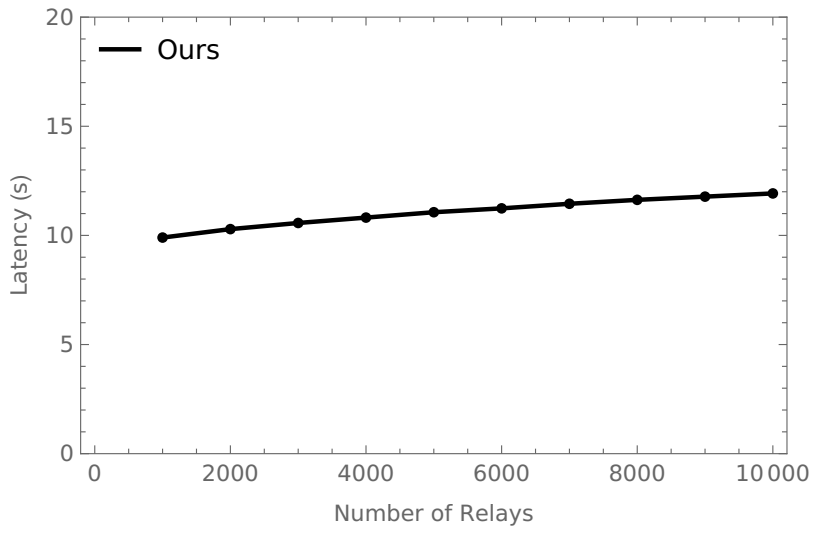

**Figure 11.** Latency of generating a consensus document under a complete DDoS attack that knocks 5 authorities offline for 5 minutes at the start of the protocol. The network returns to the normal condition of 250 Mbit/s (see Section 4.3) afterwards. The number represents the time taken to generate a consensus document after the attack ends. For comparison, the current directory protocol and the synchronous protocol take 2100 seconds (25 minutes until the next run (after the 5-minute attack ends) and 10 minutes to run the protocol) to generate a consensus document.

for a comparison of communication complexity) attributes to the issue.

In contrast, our protocol still works under the same bandwidth settings. This is because while the authorities are sending the vote document to each other in the dissemination sub-protocol, our protocol can tolerate an arbitrary latency, given that the partial synchrony model allows for it at the start of the protocol. After the dissemination sub-protocol, the message size in the agreement sub-protocol is no longer bound by the document size, allowing the protocol to keep functioning even under low bandwidth scenarios. We discover that our protocol even works under the DDoS attack condition of 0.5 Mbit/s outlined in the previous work [22], although it takes around 15 minutes to finish sending the documents and achieve consensus. We believe that this will make our protocol a strong candidate for the future Tor network for its robustness.

For completeness, we also simulate the scenario where a complete DDoS attack knocks 5 authorities offline for 5 minutes at the start of the protocol. Similar to the attack we evaluated in Section 4.3, the current and the synchronous protocol fail to generate a consensus document in this run and have to resort to the fallback mechanism that reruns the protocol after 30 minutes. In contrast, our protocol generates a consensus document in around 10 seconds after the attack ends. Figure 11 shows the latency of generating a consensus document under this scenario.

From the results of the experiment, we conclude that our protocol introduces acceptable overhead under different bandwidth settings, compared to the current insecure protocol. Notably, our protocol introduces less latency as bandwidth resources increase. For instance, when serving 9,000 relays, our protocol introduces approximately 5 seconds of additional latency at a bandwidth of 20 Mbit/s, while at a bandwidth of 50 Mbit/s, the introduced latency is reduced to approximately 3 seconds. To put it into context, the current Tor directory protocol uses a 10-minute time window to generate a consensus document. Therefore, we consider the introduced latency of several seconds to be insignificant.

**Table 1.** A comparison of designs of Tor directory protocols between the current protocol in use, the synchronous protocol proposed by Luo et al. [23], and our design. An emergency fix by Luo et al. that uses a monitor to detect the attack on the current protocol has been applied to the current Tor consensus health monitor [35].

| Protocol | Network Model | Security and Functionality | Communication Complexity |
|---|---|---|---|
| Current [37] | Bounded Synchrony | Insecure [23] (Attacks Monitored [35]) | $O(n^2d + n^2\kappa)$ |
| Synchronous [23] | Bounded Synchrony | Secure (Interactive Consistency) | $O(n^3d + n^4\kappa)$ |
| Our Work | Partial Synchrony | Secure (Interactive Consistency under Partial Synchrony) | $O(n^2d + n^4\kappa)$ |

## 7 Related Work

**Tor Directory Protocol.** The current Tor directory protocol, despite being in use for over 10 years [36], has seen very little security discussion. In 2024, Luo et al.'s work [23] gave a formal description of the protocol and pointed out its security concerns. Nevertheless, their work followed the original protocol's network assumptions in designing a new protocol. We give a comparison of the designs in Table 1. In addition, a few Tor proposals [25–28] have discussed potential improvements to the directory protocol. However, none of the proposals has been implemented yet.

**DDoS Attacks on Tor.** DDoS attacks on Tor have always been a prevalent threat. For instance, it was observed in real life that the directory authorities came under attack in 2021 [11]. A few academic works also explored the possibility. Most attacks focus on attacking the relays in the network to degrade the network's performance and potentially deanonymize users [7, 21]. Similarly, a recent study detailed an attack on the relays in 2024 [18]. An improved work by Jansen et al. [22] studied attacks on Tor bridges and scanners, and estimated that attacking these infrastructures would cost $17,000 and $2,800 per month, respectively. We find that, while the prior work focuses on the peripheral nodes of Tor and not the directory protocol itself, the brittleness of the synchrony assumption actually make Tor directory authorities a much more appealing target for a DDoS attack, since their number is low and we only need to disrupt half of them to bring Tor offline. As a result, we are able to provide a much lower cost estimate than the previous work ($53.28 per month).

**DDoS Mitigation Measures.** Many existing works study the effectiveness of different DDoS mitigation measures [33, 42, 46, 48]. While the Tor directory authorities are known to adopt some of the measures [11], the exact effect is unknown. The documentation of the DDoS prevention mechanism in Tor only lists some memory exhaustion prevention mechanisms [38], and while proof of work and rate limiting options are mentioned, they are available only for hidden services [39]. On the other hand, it is also worth noting that using commercial solutions (e.g., CloudFlare) causes the Tor network to be entirely dependent on third-party companies, which is not ideal. We think improving the protocol on the application layer to enhance its robustness is parallel to these efforts that seek to deter DDoS from lower layers, and they go hand in hand to secure the protocol.

**Network Models.** There are commonly three network models for analyzing the security of a distributed system: bounded synchrony, partial synchrony [14], and asynchrony [16]. The bounded synchronous model, as discussed in previous work [23], is used by the current Tor directory protocol [36] and states that all messages are delivered within a known time bound. These works implicitly assume perfect synchrony during execution and do not treat synchrony violation as an attack vector. The partial synchronous model relaxes the assumptions and allows the network to lose synchrony for an unknown amount of time (known as GST, the global stabilization time) before returning to a known time bound. Since in the partial synchronous model, the protocol cannot distinguish between the sender crashing or the network being slow before GST, existing work relaxes the broadcast problem's validity and only requires non-trivial output when GST is zero [1]. This paper leverages the same insight and gives a stronger validity guarantee when GST is zero. The asynchronous model, on the other hand, makes no assumptions about the network's synchrony and allows messages to be delivered in any order.

Unfortunately, the famous FLP impossibility dictates that no deterministic consensus protocol can be designed in an asynchronous network [16]. Therefore, consensus protocols under asynchronous models have to rely on randomness, most commonly in the form of a common coin [8, 9]. We observe that setting up such a system would require more advanced cryptographic primitives and stronger assumptions while also lowering the overall performance.

**Flavors of Consensus Protocols.** Historically, partial synchronous consensus protocols are designed around the ideas of views [4, 6, 47], where nodes take turns to propose a value until one of them succeeds. An emerging trend of DAG protocols allows all nodes to propose values simultaneously, and decide on the order of the values later [2, 29, 32]. We observe that such a design improves throughput at the cost of design complexity. Furthermore, adapting DAG protocols to a single-shot instance is not a trivial task, since these protocols require continuous generation of blocks to establish the leader block and hence consensus. On the other hand, the Tor directory protocol does not particularly benefit from high throughput, as it is designed to run once per hour, so each node has only one value in that hour. Therefore, we choose not to pursue this direction.

Many existing designs also seek to solve slight twists of the consensus problem. For instance, the traditionally-defined interactive consistency under bounded synchrony [15] seeks to allow every node to broadcast a value, and decide on a set that consists of these values. This is also the problem that previous works [23] tried to solve. We observe that while some works [10] claimed to have dealt with interactive consistency under other network models, the model they used is actually the same as asynchronous common subset, and not related to our work. Similar cases exist under the asynchronous model. Asynchronous common subset (ACS) [13], similar to interactive consistency, also allows every node to agree on a set of proposed values. Unfortunately, given that ACS directly implies consensus, the FLP impossibility applies, and it can only be solved with randomness. On the other hand, the deterministic gather protocol [5] allows every node to output a set that substantially overlaps with the output of other nodes, but does not guarantee agreement. Given that the directory protocol is sensitive to the input values, such a protocol will not suffice.

## 8 Conclusion

In this paper, we demonstrated a critical vulnerability in the current Tor directory protocol, highlighting its susceptibility to distributed denial-of-service (DDoS) attacks. We showed that by exploiting the bounded synchrony assumption, an adversary could disrupt the directory protocol process for as little as $53.28 per month.

To address this issue, we proposed a novel system model, interactive consistency under partial synchrony. Building on this model, we designed and implemented a secure directory protocol leveraging existing consensus protocols. Our protocol achieves resilience against DDoS attacks while maintaining performance comparable to the existing system. Through extensive evaluation, we demonstrated that our protocol introduces only minimal latency overhead, even under challenging network conditions. By separating the dissemination of relay lists from consensus generation, our approach ensures robustness under low-bandwidth and adversarial conditions. Furthermore, our findings show that adopting a partial synchrony model not only mitigates vulnerabilities but also improves the protocol's scalability and adaptability to real-world network dynamics.

Ultimately, by addressing critical vulnerabilities in the directory protocol, this research contributes to the broader goal of strengthening the resilience of anonymity networks, ensuring they remain a vital tool for preserving privacy and freedom in an increasingly adversarial digital landscape.

## Acknowledgments

We thank our shepherd, Prof. Robbert van Renesse, for his insightful comments and suggestions that significantly improved the quality of this paper.

# A Security Proof for Our Protocol

We outline a brief proof of the protocol's security over the main properties.

**Theorem A.1** (Termination)**.** *Every correct node i eventually terminates and outputs.*

*Proof.* We will prove the termination of each sub-protocol separately.

Let us start with the dissemination protocol. After *GST* passes, every correct node will have its proposal entry non-empty for every correct node. Therefore, a correct leader will receive these $(n-f)$ PROPOSAL messages and set $H_{l,i}$ to be non-empty for every correct node $i$. Since there are $(n-f)$ correct nodes, $\left|S_{l,i}\right|_{\neq\perp} \geq n-f$ and the proposal will be ready.

Termination in the agreement sub-protocol is ensured by the termination of the Byzantine agreement protocol used in the sub-protocol. Termination in the aggregation sub-protocol is guaranteed by the fact that for every node whose document appears in the output, at least one correct node has the full document. Given that the network model guarantees every message will be received eventually, the document can be retrieved as necessary. □

**Table 2.** Rounds of each sub-protocol assuming no GST. The agreement sub-protocol round complexity depends on the consensus protocol used.

| Sub-Protocol | Rounds |
|---|---|
| Dissemination | 2 |
| Agreement | Protocol-Specific |
| Aggregation | 2 |

**Theorem A.2** (Agreement). *If correct node i outputs $D_o$ and correct node j outputs $D_o'$, then $D_o = D_o'$.*

*Proof.* The consistency of the Byzantine agreement protocol in the agreement sub-protocol guarantees that every node will have the same view of the digest of every document from every node. Since all these documents can be retrieved (see Termination), every node will eventually achieve consistency by getting all the documents. □

**Theorem A.3** (Value Validity). *If $GST = 0$, for every correct node i, $V_{i,i} = D_i$.*

*Proof.* Since node $i$'s signature on the document is required for the proposal, the output $x_{i,i}$ can only be $x_i$ or $\perp$ for any correct node $i$. Specifically, when $GST = 0$, every correct node will propose a non-empty value for every other correct node, since it will have received the value within the timeout.

To propose an empty value for some node, the leader needs to show $(f + 1)$ signatures on the empty value for that node. However, since every correct node will not propose an empty value for a correct node when $GST = 0$, the leader can never receive $(f+1)$ such signatures, which is a contradiction. □

**Theorem A.4** (Common Set Validity). *For every correct node i's output $D_o$, $|D_o|_{\neq\perp} \geq n - f$.*

*Proof.* The fact that the digest vector $H$ is ready ($|H|_{\neq\perp} \geq n - f$) guarantees common set validity, since there will be at least $(n - f)$ non-empty entries. □

# B Complexity of Our Protocol

## B.1 Communication Complexity

We give a complexity analysis of each sub-protocol in our protocol. Denote $d$ as the document size and $\kappa$ as the signature size.

**Dissemination.** In the dissemination sub-protocol:

- The DOCUMENT messages are of size $O(n^2(d + \kappa))$.
- The PROPOSAL messages are of size $O(n\kappa)$ each. Therefore, the complexity is $O(n^2\kappa)$ per view, and $O((f + 1)n^2\kappa)$ in the worst-case of $(f + 1)$ views.

We conclude that the dissemination sub-protocol takes $O(n^2d + n^2\kappa)$ optimistically and $O(n^2d + n^3\kappa)$ in the worst case.

**Agreement.** The complexity of the agreement sub-protocol is determined by the Byzantine Agreement protocol that instantiates it. The input size of the protocol is $O(n^2\kappa)$.

**Aggregation.** Since the document and the signature are transmitted in the aggregation sub-protocol once, the complexity of the sub-protocol is $O(n^2d + n^2\kappa)$.

**Overall.** Adding every part together, the complexity of the protocol is

$$O(n^2d + n^3\kappa) + O(\text{Agreement}).$$

Our implementation uses HotStuff [47] as the agreement sub-protocol. Given that HotStuff over the $O(n^2\kappa)$ input has a communication complexity of $O(n^4\kappa)$, the overall complexity is $O(n^2d + n^4\kappa)$.

## B.2 Round Complexity

The round complexity analysis of the protocol is available in Table 2.

Overall, our protocol gives 4 rounds of overhead compared to the consensus protocol used. In our implementation, we use a version of improved HotStuff [17, 47] that needs 5 rounds to achieve consensus assuming a good leader and no GST, so the round complexity of our protocol will be 9 rounds with the version of HotStuff.